\documentclass[5p,twocolumn]{elsarticle}

\journal{Physics Letters B}

\usepackage{amsmath,amsfonts,amssymb,amscd,amsthm}
\usepackage{graphicx}
\usepackage{xcolor}
\usepackage{lineno}
\usepackage[figuresright]{rotating}
\usepackage{xurl}
\usepackage[breaklinks=true]{hyperref}

\biboptions{sort&compress}

\begin{document}

\begin{frontmatter}

\title{Evolution of Spectroscopic Factors in Neutron-deficient $p$-shell Nuclei in Connection with Short-Range Correlations}

\author[utokyo,ganil,riken,tohoku]{S.~Koyama\corref{cor1}}
\ead{shumpei.koyama.a4@tohoku.ac.jp}
\cortext[cor1]{Corresponding author.}

\author[utokyo,qnsi,riken]{D.~Suzuki}
\author[ganil]{O.~Sorlin}
\author[ijclab,frib,msu]{C.~Hebborn}
\author[ijclab]{M.~Assi\'e}
\author[ijclab,ganil]{L.~Lalanne}
\author[riken,cns,keio]{T.~Abe}
\author[ijclab]{D.~Beaumel}
\author[ijclab]{Y.~Blumenfeld}
\author[ganil]{L.~Caceres}
\author[ganil]{F.~De~Oliveira~Santos}
\author[lpc]{F.~Delaunay}
\author[ijclab,lpc]{F.~Flavigny}
\author[ijclab]{S.~Franchoo}
\author[lpc]{J.~Gibelin}
\author[ganil,ijclab]{V.~Girard-Alcindor}
\author[ijclab]{J.~Guillot}
\author[ijclab]{F.~Hammache}
\author[ganil]{O.~Kamalou}
\author[ganil]{A.~Kamenyero}
\author[cns]{N.~Kitamura}
\author[saclay]{V.~Lapoux}
\author[ganil]{A.~Lemasson}
\author[lpc]{A.~Matta}
\author[ganil]{B.~Mauss}
\author[ganil,dam]{P.~Morfouace}
\author[utokyo,riken]{M.~Niikura}
\author[riken]{H.~Otsu}
\author[ganil]{J.~Pancin}
\author[ganil]{T.~Roger}
\author[utokyo,cpr]{T.~Y.~Saito}
\author[utokyo,riken]{H.~Sakurai}
\author[ganil]{C.~Stodel}
\author[ganil]{J-C.~Thomas}

\address[utokyo]{Department of Physics, the University of Tokyo, 7-3-1 Hongo, Bunkyo, Tokyo 113-0033, Japan}
\address[ganil]{Grand Acc\'el\'erateur National d'Ions Lourds (GANIL), CEA/DRF-CNRS/IN2P3, Bd.~Henri Becquerel, 14076 Caen, France}
\address[riken]{RIKEN Nishina Center, 2-1, Hirosawa, Wako, Saitama 351-0198, Japan}
\address[tohoku]{Department of Physics, Tohoku University, Sendai 980-8578, Japan}
\address[qnsi]{Quark Nuclear Science Institute, the University of Tokyo, 7-3-1 Hongo, Bunkyo, Tokyo 113-0033, Japan}
\address[ijclab]{Universit\'e Paris-Saclay, CNRS/IN2P3, IJCLab, 91405 Orsay, France}
\address[frib]{Facility for Rare Isotope Beams, Michigan State University, East Lansing, Michigan 48824, USA}
\address[msu]{Department of Physics and Astronomy, Michigan State University, East Lansing, Michigan 48824, USA}
\address[cns]{Center for Nuclear Study, The University of Tokyo, Hongo, Bunkyo, Tokyo 113-0033, Japan}
\address[keio]{Quantum Computing Center, Keio University, 3-14-1 Hiyoshi, Kohoku-ku, Yokohama 223-8522, Japan}
\address[lpc]{Universit\'e de Caen Normandie, ENSICAEN, CNRS/IN2P3, LPC Caen UMR6534, F-14000 Caen, France}
\address[saclay]{CEA, Centre de Saclay, IRFU, Service de Physique Nucl\'eaire, 91191 Gif-sur-Yvette, France}
\address[dam]{CEA, DAM, DIF, F-91297 Arpajon, France}
\address[cpr]{RIKEN Cluster for Pioneering Research, 2-1, Hirosawa, Wako, Saitama 351-0198, Japan}

\begin{abstract}
We report an uncertainty-controlled determination of neutron-removal spectroscopic factors in the neutron-deficient $p$-shell nuclei 
$^{12}\mathrm{C}$, $^{10}\mathrm{C}$, $^{9}\mathrm{C}$, and $^{8}\mathrm{B}$, 
all measured under uniform conditions through exclusive ground-state-to-ground-state $(p,d)$ reactions at about 50~MeV/nucleon in a liquid hydrogen target at the GANIL/LISE facility. 
The deuterons were detected in the highly segmented MUST2 detector placed at forward angles, and the excitation energy was reconstructed by the missing-mass method.
Several transitions, including a new state in $^7$B, were observed.
Differential cross sections, which display an $L=1$ pattern for all nuclei, were fitted with 416 sets of uncertainty-quantified optical-model potential parameters and considering various single-particle bound-state wave functions to derive $C^2S$ values. 
The ratio between experimental and shell-model calculated values, $R_S$, is found to decrease slightly as a function of increasing proton-to-neutron separation energy asymmetry $\Delta S$, with a slope of $-0.0049(36)_{\text{stat}}(13)_{\text{omp}}(8)_{\text{s.p.}}$.
The observed trend is compatible with the phenomenological expectation
that the short-range correlation effect on $C^2S$ increases with the degree of nucleon minority.
The present data set delivers a high-precision benchmark that will tightly constrain future microscopic descriptions of short-range correlations in asymmetric nuclei.
\end{abstract}

\begin{keyword}
Spectroscopic factors \sep Short-range correlations \sep Transfer reactions \sep Neutron-deficient nuclei \sep Uncertainty quantification
\end{keyword}

\end{frontmatter}


\section{Introduction}
The structure of atomic nuclei is widely described by the independent-particle model~(IPM), in which proton and neutron interactions are mediated by an effective mean-field potential generated by the nuclear force~\cite{bohrmottelson}. 
Shell-model calculations add residual terms to the IPM, such as pairing or quadrupole correlations~\cite{bohrmottelson}. 
Pairing correlations induce a smoothening of the Fermi surface of a nucleus, by depleting the normally occupied orbitals to the benefit of vacant ones.
Experimentally, this Fermi surface is probed by nucleon transfer reactions~\cite{satchler}: nucleon-adding
neutron or proton transfer---such as $(d,p)$~\cite{2022KAY} and $(^3\mathrm{He},d)$, respectively---or nucleon-removal reactions such as $(p,d)$ and $(d,^3\mathrm{He})$~\cite{Flav13,Kay13,Lee10}.
It can also be probed by electron-induced proton-removal reactions $(e,e'p)$ on stable targets~\cite{Lapi93,Kramer2001}, as well as nucleon knockout reactions 
using a proton~(e.g., $(p,2p)$ or $(p,pn)$)~\cite{Atar18,Holl19} or a `heavy' ion such as Be or C~\cite{Gade08,Tost14,Tost21}, the latter being called heavy-ion knockout~(HIKO) reactions (see recent review in Ref.~\cite{2021AUMANN}).

Such experiments were first performed on stable nuclei
over a wide range of atomic masses and proton-to-neutron asymmetries, characterized by $\Delta S = S_n - S_p$ for neutron removal or $S_p - S_n$ for proton removal (up to about $\pm 10$~MeV).
These studies have found a reduction factor $R_S$ of the occupancy of the single-particle orbits of about 0.6~\cite{Kramer2001,Lapi93,Kay13,2022KAY}.
$R_S$ is obtained from the ratio of spectroscopic factors~($C^2S$) inferred from reaction measurements with respect to those predicted by the IPM or calculated through shell-model~(SM) calculations, i.e., $R_S = C^2S_{\text{exp}} / C^2S_{\text{SM}}$.
In this relation, the $C^2S_{\text{exp}}$ values are obtained by comparing experimental to theoretical cross sections, $C^2S_{\text{exp}} = \sigma_{\text{exp}} / \sigma_{\text{th}}$, with $\sigma_{\text{th}}$ obtained by means of a reaction model such as distorted-wave Born-approximation~(DWBA) calculations or more advanced calculations like coupled reaction channels~(CRCs), as explained in Refs.~\cite{satchler,thompson,timofeyuk2020}.

This reduction factor has been attributed to long-range (LRC) and short-range (SRC) nucleon--nucleon correlations~\cite{Dick04}, which are usually not included in shell-model (SM) calculations. 
Both deplete single-particle occupancies---the LRC through coupling to collective modes near the Fermi surface, and the SRC through the generation of high-momentum components. 
Electron-scattering experiments have revealed that SRCs are dominated by high-momentum neutron--proton pairs~\cite{Pias06,Sube08}. 
Furthermore, measurements on stable nuclei up to $^{208}$Pb have shown that the fraction of high-momentum components carried by the minority nucleons---the less abundant isospin species---increases with isospin asymmetry~\cite{Duer18}. 
This observation has motivated a phenomenological interpretation~\cite{2020PASCHALIS} in which the corresponding spectroscopic strengths of minority nucleons are reduced, producing a mild quenching of $R_S$ toward larger positive $\Delta S$ values.

Experimental attempts to quantify this quenching have employed a variety of reaction probes using radioactive nuclei, each providing complementary but method-dependent information on spectroscopic strengths. 
Inclusive knockout (HIKO) measurements have indicated a strong $\Delta S$ dependence and even extreme quenching for unbound final states~\cite{Gade08,Tost14,Tost21,2020CHARITY}, 
whereas transfer~\cite{Lee10,Flav13} and quasi-free scattering (QFS)~\cite{Atar18,Holl19,GomezRamos2018} studies have typically yielded larger $R_S$ values for minority species and weaker trends. 
These diverse approaches highlight the sensitivity of extracted $R_S$ values to the underlying reaction model and analysis procedure. 
At present, however, the global behavior of $R_S(\Delta S)$ cannot be established in a quantitative way, 
owing to the absence of \textit{a uniform and uncertainty-controlled methodology} that ensures consistent treatment of experimental and theoretical uncertainties.

The goal of the present study is to overcome as much as possible these limitations. 
We performed exclusive $(p,d)$ transfer reactions on $^{12,10,9}$C and $^{8}$B using identical experimental conditions, a common liquid-hydrogen (LH$_2$) target, and similar beam energies around 50~MeV/nucleon. 
All reactions remove a neutron from the same $1p_{3/2}$ orbital, enabling a direct comparison over a wide $\Delta S$ range (2.8--17.3~MeV) reaching the proton drip line for the $^{9}$C and $^{8}$B cases. 
The analysis employs uncertainty-quantified (UQ) global optical-model potentials~\cite{2023PRUITT,2023HEBBORN,Hebborn2023Err_original,Hebborn2023Err} and systematically explores model dependencies in the single-particle bound-state wave functions.

\section{Experiment}
\label{sec:sec2}
The experiment was carried out at the LISE beam line of GANIL, as reported in Ref.~\cite{2024KOYAMA}. A primary beam of $^{12}$C at 75~MeV/nucleon impinged onto a 2-mm-thick Be target to produce the $^{12}$C, $^{10}$C, and $^{9}$C nuclei in three different settings of the LISE spectrometer~\cite{lise}. 
A pure $^{12}$C beam was obtained from the primary beam, downscaled to an intensity of $1.0 \times 10^5$~pps, and slowed down in the Be target. 
The two other C isotopes, as well as $^{8}$B, were produced in fragmentation reactions and selected among
other species using a 1-mm-thick wedge-shaped Be degrader.
The last setting, tuned on $^{9}$C, included $N = 3$ isotones with fractions of $^{9}$C~(5\%), $^{8}$B~(10\%), $^{7}$Be~(65\%), and $^{6}$Li~(20\%), for a total intensity of $1.3 \times 10^5$~pps. 
A pure $^{10}$C beam was obtained due to the unbound $N = 4$ isotones of $^{8}$Be and $^{9}$B, with an intensity limited to $1.2 \times 10^5$~pps.
The $^{12}$C, $^{10}$C, $^{9}$C, and $^{8}$B beams impinged on the cryogenic LH$_2$ target~\cite{liqhydrogen}, in which the $(p,d)$ reactions occurred with comparable energies of 54.4, 55.8, 54.5, and 48.1~MeV/nucleon, respectively, at the center of the target. 
As the 6.47-$\mu$m HAVAR window foils in which the LH$_2$ was confined swelled outwards from the edge
to the center, the determination of the target thickness viewed by the beam particles could not be very precise and is estimated to be 1.5(2)~mm~(10.5(14)~mg/cm$^2$). We, however, verified that this effective thickness stayed remarkably stable (within 2\%) during the whole experiment, which is crucial for the relative studies discussed in this work.

To monitor the incident position and angle of the beam particles on target, two pairs of multi-wire proportional chambers CATS~\cite{cats}, with typical position resolutions of 1~mm~(r.m.s.), were placed upstream of the target at distances of 119 and 68~cm. 
The time difference of the CATS signal with respect to the radio-frequency signal of the cyclotrons was used to identify and select the $^{9}$C and $^{8}$B nuclei on an event-by-event basis among other transmitted nuclei. 

An array of six telescopes of the charged-particle MUST2 detector was used to detect recoiling deuterons from $2^\circ$ to $36^\circ$ in the laboratory, as described in Fig.~1 of Ref.~\cite{2021LAL}.
Each telescope consists of a double-sided silicon strip detector~(DSSD) followed by CsI(Tl) detectors. 
The DSSD has an active area of $98 \times 98$~mm$^2$ with a thickness of 300~$\mu$m.
Each detector is segmented into 128 strips with a 0.76~mm pitch.
The energy resolution was about 40~keV FWHM for 5.5-MeV $\alpha$ particles of an $^{241}$Am standard source. 
The total kinetic energy of the deuterons was deduced from the sum of the energies measured by the DSSD and the CsI(Tl) detectors. 
The particle identification was performed by the $E$--$\Delta E$ method. 
The scattering angle of a recoiling particle was obtained from the hit position measured by the DSSDs, after correcting for the position and angle of the incident ion on target.

\begin{figure}[t]
 \centering
 \IfFileExists{./fig1-crop.pdf}{%
   \includegraphics[width=\columnwidth]{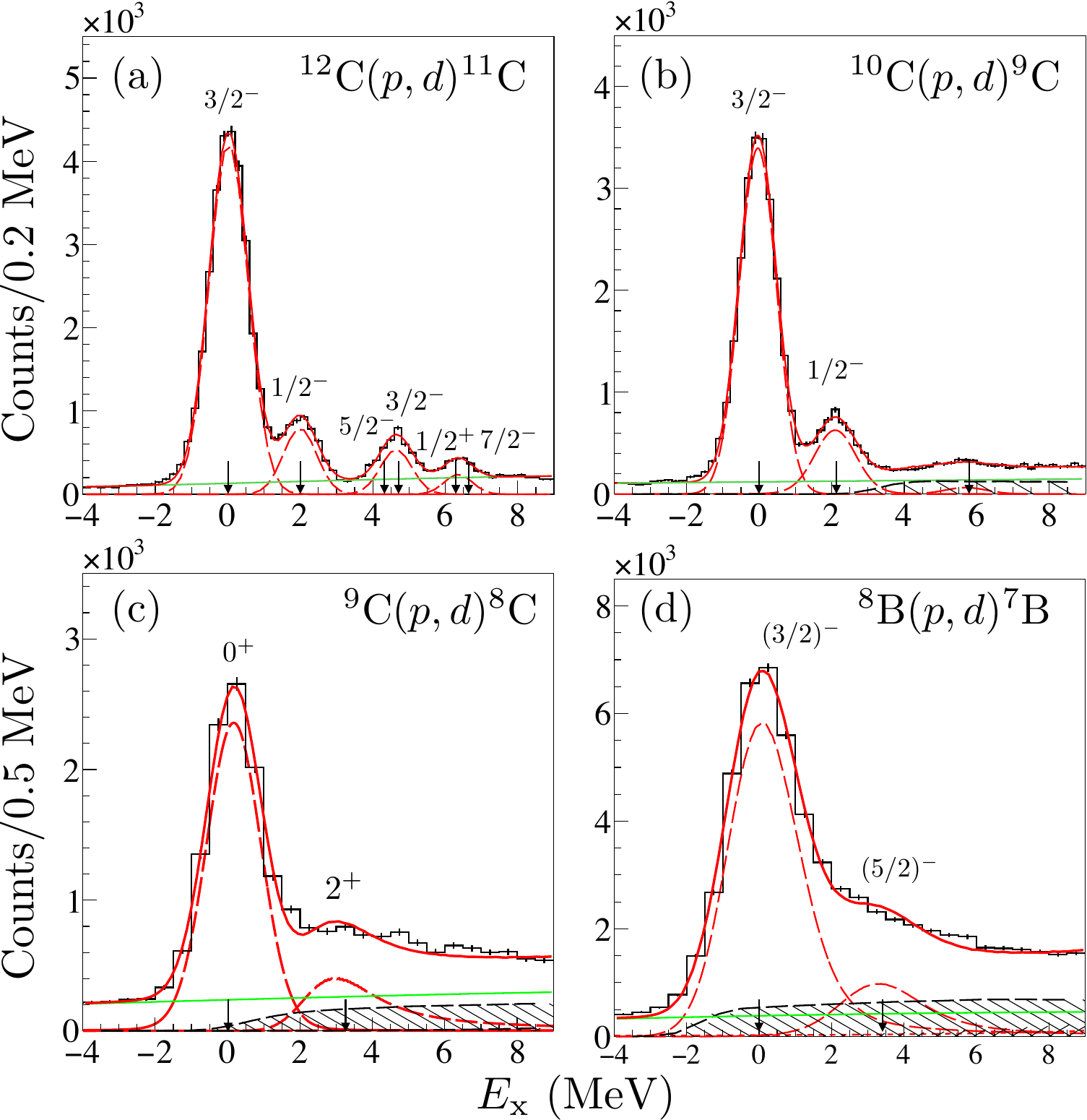}%
 }{%
   \framebox{\parbox{0.9\columnwidth}{\centering\vspace{1.8cm}\textbf{Figure 1 Placeholder}\\\small\textit{fig1-crop.pdf (Excitation energy spectra)}\vspace{1.8cm}}}%
 }
 \caption{\label{fig:fig1} 
The excitation energy spectra of $^{11}$C, $^9$C, $^8$C, and $^7$B from the $(p,d)$ reaction.
The results of the fit are also shown for the total fit~(red solid curve) and each state~(red dashed curve). 
The background (in green) is estimated from the data taken without liquid hydrogen. 
For the $^{9}$C, $^{8}$C, and $^7$B spectra, non-resonant components are shown by the shaded area. 
Excitation energies of the states, labeled with their spins and parities, are indicated by arrows.
The excited states above 3~MeV in $^{11}$C are not fully separated but are known to be populated in previous studies~\cite{1980COMFORT,1980HOSONO}.}
\end{figure}

\section{Analysis and results}
\label{sec:sec3}
The excitation energies and the center-of-mass scattering angles~($\theta_{\text{c.m.}}$) of $^{11}$C, $^9$C, $^8$C, and $^7$B were reconstructed by means of the missing-mass method, using the energy and angle of the recoiling deuteron.
The angle-integrated excitation energy spectra up to $\theta_{\text{c.m.}} = 45^\circ$ are shown in Fig.~\ref{fig:fig1} up to 9~MeV.
The background from the target cell was estimated by fitting a third-order polynomial function to the spectrum taken without liquid hydrogen.
The obtained background distribution was scaled by the number of incoming beam particles and is shown by a green curve, which well reproduces the excitation energy below 0~MeV, where no peak is present.

The spectra are fitted with symmetric Gaussian functions~(for bound states) or the convolution of a Gaussian and a Lorentzian~(for unbound states) in order to obtain the yield of each final state.
The width of the Gaussian functions~(0.5~MeV r.m.s.) is due to the geometrical and intrinsic resolutions of the CATS and MUST2 detectors, as well as to the uncertainty of the energy loss of the particles in the target, which is well reproduced by the NPTOOL simulation~\cite{2016MAT}. 
The details of the fit to the $^8$C spectrum, which includes the broad first $2^+$ resonance near the ground state~(g.s.) as well as non-resonant components, have been described in Ref.~\cite{2024KOYAMA}.
The fits of the $^9$C and $^7$B spectra have been performed using the same procedure.
In addition to the known states, a new one is observed in $^7$B at 3.2(2)~MeV, with a decay width of 3.0(4)~MeV. 
This state is likely the mirror of the $5/2^-$ state at 2.9(3)~MeV in $^7$He~\cite{1999KOR}. 
For all spectra, the uncertainty from the fit~(mostly due to the subtraction of background components) is the smallest for the g.s. peak. Consequently, we will consider $(p,d)$ transfer reactions of g.s.-to-g.s. transitions~($T_{gg}$) in the present work, although others could also be determined.
\begin{table*}[t]
  \centering
  \caption{Summary of the neutron-removal experimental $C^2S_{\text{exp}}$ and calculated $C^2S_{\text{SM}}$ values for g.s. ($J^\pi_i$) to g.s. ($J^\pi_f$) transitions $T_{gg}$ for all isotopes with various $\Delta S$ (in MeV), as well as their corresponding reduction factors $R_S$.
  $C^2S_{\text{SM}}$ values are obtained from the average of the shell-model calculations with two interactions. 
  The uncertainties associated with the fit of experimental transfer data using the KDUQ representative parameter set~(statistical), the optical model, and the s.p. binding potential are shown respectively in the first, second, and third parentheses for $C^2S_{\text{exp}}$ and $R_S$ (see text for details).}
  \label{tab:single_config}
  \begin{tabular}{c|cccccc} \hline\hline
             & $J_i^{\pi}$ & $J_f^{\pi}$ & $\Delta S$ (MeV) & $C^2S_{\text{exp}}$ & $C^2S_{\text{SM}}$ & $R_S$ \\ \hline
    $^{12}$C & $0^+$       & $3/2^-$     &  2.8             & 1.70(11)(12)(19)    & 2.82(3)            & 0.60(4)(4)(7) \\
    $^{10}$C & $0^+$       & $3/2^-$     & 17.3             & 0.92(6)(7)(11)      & 1.74(1)            & 0.53(4)(4)(6) \\
    $^{9}$C  & $3/2^-$     & $0^+$       & 12.9             & 0.50(5)(3)(8)       & 0.87(1)            & 0.58(5)(3)(5) \\
    $^{8}$B  & $2^+$       & $(3/2)^-$   & 12.7             & 0.50(6)(3)(8)       & 0.82(5)            & 0.61(7)(4)(5) \\ \hline\hline
  \end{tabular}
\end{table*}

\begin{figure}[t]
 \centering
 \IfFileExists{./fig2-crop.pdf}{%
   \includegraphics[width=\columnwidth]{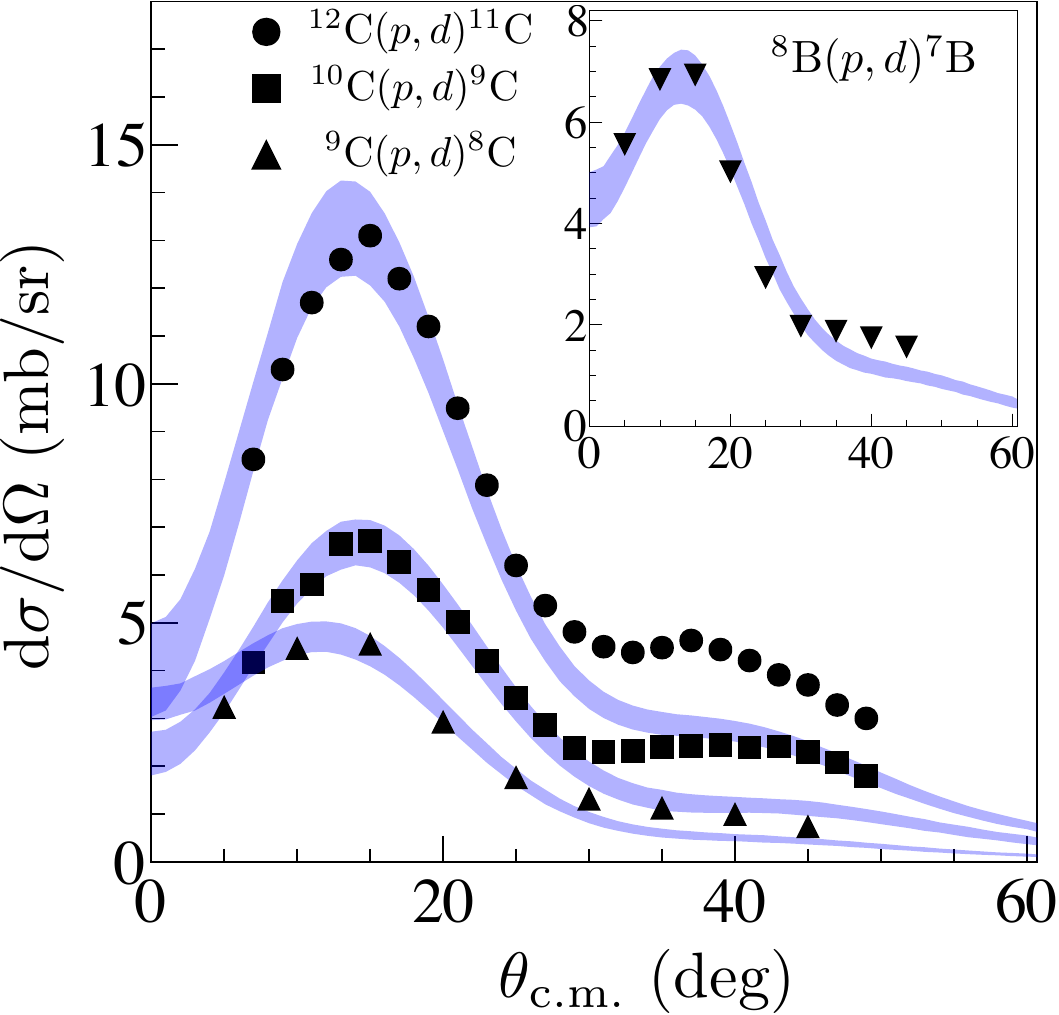}%
 }{%
   \framebox{\parbox{0.9\columnwidth}{\centering\vspace{1.8cm}\textbf{Figure 2 Placeholder}\\\small\textit{fig2-crop.pdf (Differential cross sections)}\vspace{1.8cm}}}%
 }
 \caption{\label{fig:fig2} 
   Differential cross sections of $T_{gg}$ of the $^{12}$C$(p,d)^{11}$C, $^{10}$C$(p,d)^{9}$C, and $^{9}$C$(p,d)^{8}$C reactions~($^{8}$B$(p,d)^{7}$B in inset).
   The uncertainty from the global OMP is shown by the blue band~(obtained with a fixed $r_0=1.33$~fm).
 }
\end{figure}

The experimental differential cross sections for $T_{gg}$ in Fig.~\ref{fig:fig2} have been obtained for each isotope of interest by fitting their excitation energy spectra of Fig.~\ref{fig:fig1} for several angular bins. The geometrical acceptance of the detectors was estimated by the NPTOOL simulation~\cite{2016MAT}. Their amplitude for the C isotopes displays an almost gradual reduction in concert with the decreasing number of neutrons in the $1p_{3/2}$ orbital, from four in $^{12}$C to only one in $^{9}$C. These distributions were fitted with DWBA calculations~\cite{satchler,thompson,timofeyuk2020} using the DWUCK5 code to extract spectroscopic factor values $C^2S_{\text{exp}}$. The adiabatic approximation was applied to the exit channel to treat the deuteron breakup, constructing the deuteron--nucleus potential from nucleon--nucleus potentials~(see Eq.~(5) of Ref.~\cite{adiabatic}).
While CRC calculations can in principle provide a more complete treatment, previous benchmark studies have shown that CRC effects modify extracted spectroscopic factors by at most 5\% for $(p,d)$ reactions at comparable energies~\cite{Del2005}. Therefore, a single-step DWBA analysis was adopted as a sufficiently accurate and transparent framework for the present systematic study.
In the present work, all potentials for both entrance and exit channels are derived from KDUQ global nucleon potentials~\cite{2023PRUITT}, which use the same parametrization and corpus of data as the KD03 potential~\cite{KONING2003}. 
The various s.p. bound-state wave functions of a neutron in the potential of the $(A-1)$ nucleus
were calculated for the $(p,d)$ reactions by using Woods--Saxon potentials with radii within the $1\sigma$ interval $r_0=1.33(5)$~fm, taken from the average of Hartree--Fock (HF) calculations~\cite{Hai2024}, a diffuseness parameter of 0.65~fm, and a depth fitted to reproduce the neutron separation energy. 
This adopted value of 1.33(5)~fm is consistent with those used in previous analyses of the quenching of spectroscopic factors~\cite{Lapi93,Kay13,2022KAY,2023HEBBORN,Hebborn2023Err_original,Hebborn2023Err}. It is also consistent with $r_0 = 1.35$~fm used for the $^{12}$C$(e,e'p)^{11}$B$_{\text{g.s.}}$ reaction~\cite{Kramer2001}, which is the mirror of the $^{12}$C$(p,d)^{11}$C$_{\text{g.s.}}$ reaction we are studying.
Fig.~\ref{fig:fig2} shows that the experimental angular distributions from the $(p,d)$ reaction are accurately reproduced by the $1\sigma$ theoretical band using an $L = 1$ transferred angular momentum from the $1p_{3/2}$ orbital. It has been obtained from the 416 sets of the KDUQ posterior distributions~\cite{2023PRUITT} by fixing $r_0=1.33$~fm. 

\begin{figure}[t]
 \centering
 \IfFileExists{./fig3_simple-crop.pdf}{%
   \includegraphics[width=\columnwidth]{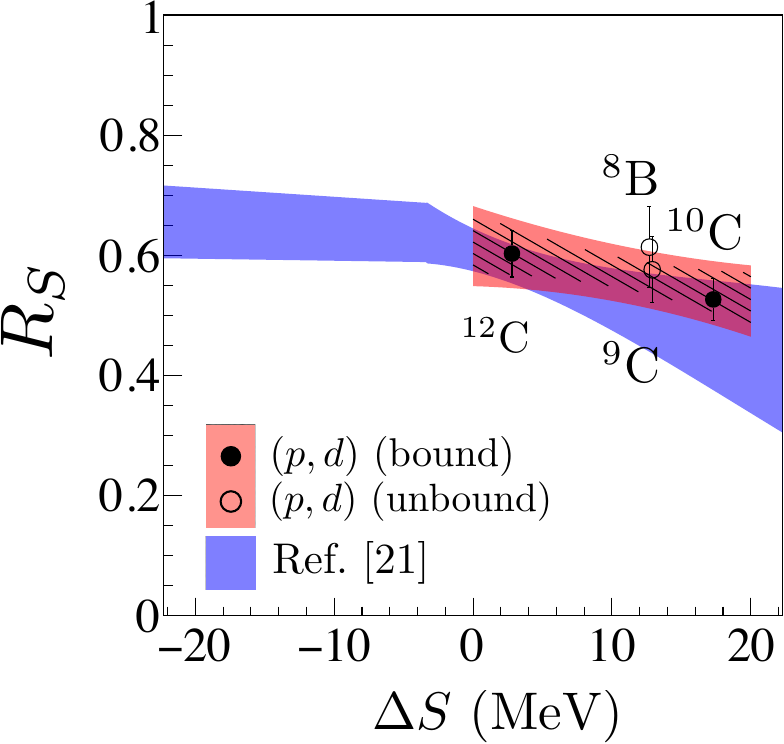}%
 }{%
   \framebox{\parbox{0.9\columnwidth}{\centering\vspace{1.8cm}\textbf{Figure 3 Placeholder}\\\small\textit{fig3\_simple-crop.pdf (Reduction factor $R_S$ vs $\Delta S$)}\vspace{1.8cm}}}%
 }
 \caption{\label{fig:fig3} 
   Reduction factor from the experimental results of the current study with the $(p,d)$ reaction for the g.s.-to-g.s. transitions $T_{gg}$ in the $^{12,10,9}$C and $^{8}$B nuclei. Filled (open) circles correspond to cases leading to bound (unbound) final states. The hatched area shows the statistical uncertainty only, while the red band also includes the systematic uncertainty from the choice of optical-model potentials. The blue band corresponds to the phenomenological model of Ref.~\cite{2020PASCHALIS}.
 }
\end{figure}

To determine $C^2S_{\text{exp}}$ for the four studied nuclei, the experimental differential cross sections are compared in Fig.~\ref{fig:fig2}, up to an angular range of $25^\circ$, to the median values of the KDUQ posterior distributions with $r_0=1.33$~fm~(referred to later in the text as the KDUQ representative parameter set). 
Our experimental $C^2S_{\text{exp}}$ value of 1.99(11) for $T_{gg}$ in $^{12}$C compares reasonably well (given the uncertainty on our LH$_2$ target thickness) with the mean value of 1.70 obtained from the results of two previous $^{12}$C$(p,d)^{11}$C$_{\text{g.s.}}$ experiments~\cite{1980COMFORT,1980HOSONO}, reanalyzed using the KDUQ representative parameter set.
In the following, we normalize the $T_{gg}$ value of $^{12}$C to 1.70, thus not considering the uncertainty related to the target thickness. 
This is justified by the fact that we are interested in the evolution of $C^2S_{\text{exp}}$ values, which would all be affected by the same amount if the target thickness is changed. 
The obtained values are given in Table~\ref{tab:single_config}, together with their systematic uncertainties resulting from the choices of OMP parameters and $r_0$ values. 

The $C^2S_{\text{exp}}$ values for all studied nuclei are then compared in Table~\ref{tab:single_config} to those obtained from SM calculations, $C^2S_{\text{SM}}$, in order to derive the corresponding reduction factor, $R_S$, and its evolution towards the drip line.
SM calculations were carried out by the KSHELL~\cite{kshell} code using the CK~\cite{cohen_kurath} and the YSOX~\cite{ysox} interactions.
For all cases, the largest $C^2S$ values by far ($>70\%$) correspond to $T_{gg}$.
Moreover, such calculated values with the two interactions differ by less than $\pm 2\%$ for all C isotopes and $\pm 6\%$ for $^8$B. Average values of the SM calculations with the two interactions and uncertainties obtained from their difference were considered to determine $R_S$ in Table~\ref{tab:single_config}.

The evolution of $R_S$ as a function of $\Delta S$ is fitted with a linear function in Fig.~\ref{fig:fig3}. The error bars of the data points include statistical uncertainties arising from the fits of the angular distributions with the KDUQ representative parameter set. The OMP uncertainties of the slope and intercept parameters were obtained from 416 linear regressions of $R_S$ extracted from each reaction data set using a fixed $r_0=1.33$~fm, for all 416 KDUQ samples, hence accounting for their correlations. Moreover, we quantify the uncertainties of the linear regressions using the average of the covariance matrices associated with the 416 fits. Finally, we evaluated the uncertainties from the choice of the $r_0$ parameter in the s.p. binding potential. To do so, the previous analyses were repeated by varying $r_0$ within the standard deviation of $\pm 0.05$~fm found in HF calculations for light nuclei~\cite{Hai2024}, using a common $r_0$ value for all nuclei in each iteration. We obtain $R_S = -0.0049(36)_{\text{stat}}(13)_{\text{omp}}(8)_{\text{s.p.}}\,\Delta S + 0.62(5)_{\text{stat}}(5)_{\text{omp}}(6)_{\text{s.p.}}$. In Fig.~\ref{fig:fig3}, the statistical uncertainty is shown by the hatched zone and the one adding the OMP uncertainties is indicated by the red band. It is important for the following discussion to notice that the effect of $r_0$ on the slope value is small. Variations in $r_0$ influence mostly the overall normalization of the extracted $C^2S$, hence leading to a large uncertainty of the intercept value.

\section{Discussion}
\label{sec:sec4}
The experimental red band of Fig.~\ref{fig:fig3} shows a weak dependence of $R_S$ with $\Delta S$, even when including the neutron-removal cases from $^{9}$C and $^{8}$B leading to the unbound $^{8}$C and $^{7}$B nuclei, respectively. This is at variance with the steep decreasing slope obtained from all HIKO results~\cite{Gade08,Tost14,Tost21}, and with the even weaker $R_S$ value of about 0.15 for the neutron removal from $^{9}$C~\cite{2020CHARITY}. It is clear from our work that no significant additional quenching is found for the $^{9}$C and $^{8}$B nuclei at the verge of the proton drip line, where $R_S \simeq 0.6$ for both cases (see Table~\ref{tab:single_config}). Discrepancies in $C^2S$ values inferred from HIKO and transfer data have already been observed in previous works (see review in Ref.~\cite{2021AUMANN} and references therein), which suggests that they are due to an inaccurate theoretical description of the HIKO reaction (see more extensive discussion in Refs.~\cite{2021AUMANN,Hebborn2023Err_original,Hebborn2023Err}). 

The conclusion of a mild dependence of $R_S$ with $\Delta S$ agrees with that already derived from other transfer reactions~\cite{Flav13,Kay13,2022KAY,Lee10} or QFS experiments~\cite{Atar18,Holl19,GomezRamos2018}. Indeed, these experiments were all compatible with no dependence of $R_S$ with $\Delta S$. However, most of them could not be conclusive owing to their rather large systematic uncertainties.

From our work, we obtain a downsloping trend of 
$-0.0049(36)_{\text{stat}}(13)_{\text{omp}}(8)_{\text{s.p.}}$. 
While this slightly decreasing trend is statistically compatible with zero, our high-precision data set constrains the $\Delta S$ dependence to be mild.
This result is consistent with the slope obtained from a reanalysis of 
$(p,d)$ transfer reactions on Ar isotopes~\cite{Lee10,2023HEBBORN,Hebborn2023Err_original,Hebborn2023Err}. 
In the present work, the optical-model uncertainty is evaluated by explicitly 
accounting for correlations among the KDUQ samples through multiple correlated 
linear regressions, leading to a reduced and more robust estimate of the 
``omp'' systematic uncertainty. Moreover, the present analysis includes an 
evaluation of the ``s.p.'' uncertainty. Together, our work further extends the uncertainty 
treatment beyond the previous studies.

The blue band shown in Fig.~\ref{fig:fig3} corresponds to the phenomenological 
model of Ref.~\cite{2020PASCHALIS}, which was derived from fits to $(e,e'p)$ 
experimental data on (magic) stable nuclei as a function of $(N-Z)/A$ 
(later converted to $\Delta S$). 
This parametrization assumes an enhanced involvement of minority nucleons 
in short-range-correlated ($n$--$p$) pairs~\cite{Duer18}. 
Applied to the neutron-deficient isotopes studied here, this leads to a reduced 
probability for neutron removal, and hence to a decrease of $R_S$, as minority 
neutrons are increasingly involved in high-momentum SRC pairs. 
The model thus predicts two distinct regimes: an almost constant behavior of 
$R_S$ for negative $\Delta S$ values and a shallow downsloping trend for positive 
$\Delta S$ values.

The present data set provides a sensitive test of the positive-$\Delta S$ 
region with a wide $\Delta S$ coverage. While previous studies~\cite{Lee10,2023HEBBORN,Hebborn2023Err_original,Hebborn2023Err} span both positive 
and negative $\Delta S$ values, their coverage on the positive side 
was more limited. 
By focusing on the positive-$\Delta S$ region, the present work specifically 
probes the regime where a decreasing behavior of $R_S$ is predicted. 
The observed downsloping trend is consistent with this phenomenological 
expectation, and our extracted value of $R_S$ at $\Delta S = 0$, 
$0.62(5)_{\text{stat}}(5)_{\text{omp}}(6)_{\text{s.p.}}$, agrees well with the 
value of 0.61(4) predicted by Ref.~\cite{2020PASCHALIS}.

This agreement with the phenomenological model may support the role of SRC in 
inducing an additional quenching of spectroscopic strengths for minority 
species. However, the magnitude of the effect remains small, and firm 
conclusions will require theoretical frameworks that treat nuclear structure 
and reaction dynamics on the same footing, including short- and long-range 
correlations as well as coupling to the continuum. 
In this regard, the present results provide a high-precision benchmark that 
should stimulate and constrain such future theoretical developments.

\section{Conclusions}
\label{sec:sec5}
The $(p,d)$ transfer reactions using $^{12}$C, $^{10}$C, $^9$C, and $^8$B beams were studied at about 50~MeV/nucleon,
with the goal of probing the evolution of neutron-removal spectroscopic factors in nuclei with increasingly large proton-to-neutron imbalance up to $(Z-N)/A=0.33$, even reaching the proton drip line.
Many transitions, including a new $^7$B state at 3.2(2)~MeV, were observed, and g.s.-to-g.s. transitions were used to derive this evolution. In the present work we have minimized systematic uncertainties for all nuclei by using the same experimental conditions, the same optical model, and considering reactions removing a neutron from the same s.p. orbital. The resulting data set provides an \textit{internally consistent and uncertainty-controlled transfer-reaction benchmark} for probing the evolution of spectroscopic strengths toward asymmetric systems. We find that the ratio between experimental and theoretical spectroscopic factors, $R_S$, smoothly decreases with $\Delta S$.
We do not observe additional quenching for $^9$C and $^8$B at the drip line, as in Ref.~\cite{2020CHARITY}. The slope of decrease of $R_S$ with $\Delta S$ agrees with the one obtained in Ref.~\cite{2020PASCHALIS}, which considers that minority nucleons are more involved in SRCs. The present data set also intends to stimulate the development of theoretical frameworks treating on an equal footing the scattering dynamics and bound or unbound state properties to study the interplay between nuclear structure and short-range correlations in nuclei with large neutron-to-proton imbalance.

\section*{Acknowledgements}
The authors thank the accelerator staff and the technical staff of GANIL.
We thank the ECT* center of theoretical physics as the idea of this work emerged through discussions in a meeting dedicated to the physics at the drip lines. 
We thank S. Paschalis and D. Y. Pang for useful discussions. 
S. K. was supported by the JSPS Grant-in-Aid for JSPS Research Fellows JP16J04726.
This work was supported by the JSPS KAKENHI Grant Numbers JP19H01914, JP21K03564, and JP24H00239.
This work was supported by the RIKEN TRIP initiative (Nuclear transmutation).

\bibliographystyle{elsarticle-num}
\bibliography{c2s}

\end{document}